\documentclass{iopart}
\usepackage{iopams,bm,epsfig}
\usepackage{graphicx}
\usepackage{exscale}
\usepackage{subfigure}
\usepackage{xspace}
\newcommand{\ket}[1]{\left| #1 \right\rangle}
\newcommand{\bra}[1]{\left\langle #1 \right|}

\newcommand{\be}{\begin{equation}}
\newcommand{\ee}{\end{equation}}
\newcommand{\bea}{\begin{eqnarray}}
\newcommand{\eea}{\end{eqnarray}}
\newcommand{\edoc}{\end{document}}

\graphicspath{%
    {converted_graphics/}
    {/}
    {Revise2/}
}
\begin{document}

\title{ Exceptional points and quantum correlations in precise measurements }

\author{ A. Thilagam} 
\address{Information Technology, Engineering and the  Environment, 
Mawson Institute,
University of South Australia, Australia
 5095}
\date{\today}
\begin{abstract}
We examine  the  physical manifestations of exceptional points and passage times 
in a two-level system which is subjected to quantum measurements and 
which  admits a non-Hermitian description. Using an
 effective Hamiltonian acting in the two-dimensional space
 spanned by the evolving  initial and final states, the effects 
of  highly precise  quantum measurements in which the monitoring device interferes significantly with the 
evolution dynamics of the monitored two-level system is  analysed.
The  dynamics of a multipartite  system consisting of the  two-level system, a source 
of external potential   and the measurement device is examined using correlation
 measures such as entanglement and  non-classical quantum correlations. 
Results show that the  quantum correlations between the monitored (monitoring) systems  is considerably decreased (increased) as the measurement precision  nears the exceptional point, at which the 
passage time is half of the measurement duration. The results indicate that
the underlying mechanism by  which the non-classical correlations of quantum systems 
are transferred from one subsystem to another may be better revealed via  use of  geometric approaches.

 \end{abstract}
\maketitle

\section{Introduction}\label{intro}

Non-Hermitian systems  \cite{sense,benderB,rotter,nest09,bend,fring,ali,mosta,bendJ,heissN}
play important roles in the dynamics of open quantum systems, and the appearance of non-Hermitian 
 terms   have profound 
 implications for various physical, chemical and biological systems
\cite{heissN,hern,muss,jin,chem,berry,micro2011,exptex,Goet,lee,thilchem}.
The striking difference between  non-Hermitian physics and  Hermitian physics lies in the
occurrence of degeneracies such as  exceptional points. 
This is the case even in situations where the non-Hermitian system 
is almost similar to its  Hermitian counterpart. 

The exceptional point is a topological defect which occurs when 
 two eigenvalues of an operator coalesce
as a result of a change in selected system parameters, with  the
 two mutually orthogonal  states merging into one self-orthogonal state, 
 resulting in a singularity in the spectrum \cite{Heiss}. 
The critical parameter values  at which the singularity appear  are called exceptional
points. These  points are known to be  located  in the vicinity
of a level repulsion \cite{Heiss} and unlike degenerate points,
only one eigenfunction exists at the exceptional point due to the merging of
two eigenvalues. This implies that the branch point at which the exceptional
point forms is no longer single valued.  The Berry phase around the exceptional
point is generally treated using conventional techniques \cite{Anand} which involves tracking
a closed path in the Riemann space, however the procedure here involves transversing
the loop twice since the eigenstates swaps positions after the  first loop.
From a quantum information perspective, the emergence of the
exceptional point can be seen as a loss in information due to the decrease
of eigenvalues in the parameter space, and it would be worthwhile
to examine whether such a loss is counteracted by changes in another measurement
subspace. In a recent work, we identified the critical temperatures at which exceptional points
occurs in photosynthetic systems \cite{thilchem}. The
 experimental detection of the exceptional  points remains a challenge, even though the
topological properties of the singularities at exceptional points are accessible
to experimental work, which enables  crossings
and anticrossings of energies and widths to be observed \cite{exptex}.

Other than the appearance of exceptional points, quantum systems with non-Hermitian components
 evolve in vastly different ways  from  systems 
of a   purely Hermitian Hamiltonian. Unlike the long lifetimes of states associated with an Hermitian Hamiltonian,
  those of the  non-Hermitian Hamiltonian
have a finite lifetime. Moreover  states which are orthogonal  under the ordinary 
inner product in the Hermitian quantum space,  exist under non-orthogonal forms
in the  non-Hermitian case. There are also differences in the context of the
 quantum brachistochrone problem.
 In the general brachistochrone problem, the minimum time taken to transverse the 
path  between two locations of a particle under a set of constraints is to be determined.
This problem translates to the quantum brachistochrone case when  extended to  the evolution of quantum states.
It has been shown that the passage time of  evolution of an
initial state  into the final state can be made arbitrarily small for
a time-evolution operator  that is  non-Hermitian but {\it PT}-symmetric \cite{bendJ}. This result
has recently been generalized  to  non-{\it PT}-symmetric dissipative
systems \cite{fring}, with the results in Refs. \cite{fring,bendJ}  indicating that 
the time scales of propagation in non-Hermitian quantum mechanics  are faster than those
of  Hermitian systems. In a recent work, Nesterov \cite{nest09,nestsigma}
examined features of the non-Hermitian Hamiltonian and the associated 
quantum brachistochrone problem using a suitable  geometric basis
based on the  Fubini--Study metric on the complex Bloch sphere.

In this work, we examine  the  physical manifestations of systems which are subjected to 
quantum measurements, and which  admit an effective non-Hermitian description.
The association between the non-Hermitian dynamics of open quantum systems and 
quantum measurements has interesting consequences not explored in earlier works. 
Firstly,   the appearance of degeneracies such as 
 exceptional points, can be extended to systems undergoing quantum measurements.
Secondly, the   quantum non-Hermitian brachistochrone problem \cite{ali,bendJ}
can be analyzed in the context of quantum measurements. 
Investigations of quantum measurements in 
a range of correlated quantum systems  highlight
the distinct differences between the quantum and  classical 
worlds  \cite{von,zeh,ob,Misra,bus,wise,It,FacJ,brakha,menbk,zuro,schloss,joh}.  
To this end, it would be interesting to seek further understanding of quantum systems which are monitored
by an external observers using the non-Hermitian description of the measured system. These issues form the key
motivation for the current study.

Quantum measurements   can be viewed 
as interactive processes which have positive or negative outcomes
 in the detection of the observed entity. The act of a measurement has  a wide 
ranging degree of influence on the system under observation, and  is 
dependent on the level of precision with which the measurement is made.
The assumption that measurement has no influence on the monitored system
would enable the determination of observables via consecutive measurements
based on the first reference state. This would imply  the  violation
 of  the quantum uncertainty of non-commuting variables. In this 
regard, quantum measurements can be seen to facilitate
the creation of information linked to an observation. Accordingly 
 the act of measurement is irreducible within the framework of quantum
mechanics,  unlike  other types of interactions. This obvious discrepancy between
the dynamical evolutions at the microscopic level and outcomes at the
macroscopic levels  highlights the  inconsistencies  of  quantum measurements.
The unitary and reversible features of the Schr\"odinger
equation and  the non-unitary elements inherent in the projection
postulate are clearly incompatible. However both these core
 processes need to be unified in order  to examine the influence of a continuous monitoring  
on the evolution of a quantum system, which is indeed a challenging
task. 

 While all kinds (low or high precision)  measurements
invariably disturb the measured system, there are 
differences in the degree of these disturbances.
In the case of  low precision  measurements,
a device $D$ introduces  minimal disturbance on the measured system, 
$S$ with state $u|S_u\rangle + v|S_v\rangle$. 
The state of the measuring device can be $|D_u\rangle$ or 
$D_v\rangle$ after the measurement, and  is different from
its state before measurement, $|D_i\rangle$. The composite
system $S \otimes D$ proceeds in an  approximately unitary 
fashion as $U |S_u\rangle |D_i\rangle = |S_u\rangle |D_u\rangle$,
$U |S_v\rangle |D_i\rangle = |S_v\rangle |D_v\rangle$.
In the case of ideal measurements, the resulting state of the
system after measurement  generally belongs to the set
of the orthonormal basis of the quantum system. In both
 ideal or weak measurements, the observables manifests as  Hermitian operators 
 which act on the state space, and  the eigenvalues of the
eigenstates at which the system existed during the period of  measurement
can be obtained.

In the case of high precision  measurements, the measuring
device  subjects the measured system to significant  disturbances,
and further  analysis of the quantum evolution 
becomes complicated as  the systems tracks
complex routes, possibly  due to the appearances
of non-Hermitian terms. In general,  it is  difficult to determine
 the intended reading  or the state that the monitored system is
 actually existing in  after the measurement
process. To investigate some of these issues, we employ
an approach  based on  the non-Hermitian dynamics of a two-level 
system which was originally solved with the aim of seeking a link between a decay term 
and Berry's phases by Garrison and Wright \cite{gar}.
We associate an analogous decay  term to level of  
precision to the  quantum measurement problem, so that a higher  measurement precision results
in a larger magnitude of this decay term. 

Our paper is organized as follows. In Section \ref{feyn} we provide 
a brief review  of  the  Feynman's path integral framework
which  admits an effective non-Hermitian description for quantum measurements.
The appearance of exceptional points during quantum measurements
 in the two-dimensional space  of a simple two-level  system is examined in Section \ref{excepts},
explicit expressions of the passage times are also provided in this Section. To further understand the
 dynamics of non-classical correlations, particularly near the exceptional point, we
consider a  multipartite  system consisting of the  two-level system,  a potential source  and the 
 monitoring device in  Section \ref{disco}. The dynamics of this multipartite 
system  is examined using well known  correlation measures such as Wooters concurrence and
the quantum correlations in  Section \ref{disco}. The conclusions are  presented in Section \ref{con}.

\section{Non-Hermitian features of quantum measurements}\label{feyn}

Our starting point for the non-Hermitian analysis of quantum measurements  is the
 the Feynman's path integral framework \cite{Feyn1,Feyn2}. 
in which  the probability
amplitude of  transitions from the initial to the final state of the system
is obtained via  summation of the  amplitudes of  all  
possible  paths.  A  weight entity which provides a measure of contribution of each constituent path.
is associated with the individual paths which are involved in the summation.
In this work,  we employ  a variant of the Feynman's path integral
formalism based  on the restricted path integral approach \cite{menbk,mensky}. 
In the restricted path approach,  the  continuous measurement 
of a quantity with a given  result is monitored by 
constraints imposed on the Feynman's path integral. 

The restricted path integral is derived \cite{menbk,mensky,menprl,men}
from the Feynman path integral  through the incorporation
of  a weight functional within an integrand which incorporates 
 the various paths involved 
in the summation process. 
We recall that the Feynman's propagator, 
$K_{[E]}(q^{\prime},\tau;q,0)$ in the phase-space representation
 at time $\tau$ is given  by \cite{Feyn1,Feyn2}
\be
K(q^{\prime},\tau;q,0)=
\int d[q] d[p] e^{ 
{i \over {\hbar}}\int_0^{\tau}[p \dot{q} - \widehat H_0(q,p)]dt }
\label{fey1}
\ee
where $\widehat H_0$ is the Hamiltonian of
the closed (unmeasured) quantum system and $[p]$ and $[q]$ are the
 paths in the momentum  and 
configuration spaces respectively. In Mensky's formalism,
 the output of a quantum system subjected to measurement
is expressed in terms of constrained paths linked  to
the measured system via  a  weight functional $w_{[E]}$ \cite{menbk,mensky,menprl,men}. 
The  functional may assume  a  Gaussian form with a
damping magnitude that is proportional to the 
squared difference between the observed value along  
the paths and the actual  measurement result. 
 Thus a system subjected to measurement evolves via a propagator
which modifies  Eq.(\ref{fey1}) according to \cite{men,ono1}
\be
K_{[E]}(q^{\prime},\tau;q,0)=
\int d[q] d[p] e^{ \frac{i}{\hbar}
\int_0^{\tau}[p \dot{q} -\widehat H_0(q,p)]dt }
w_{[E]}
\label{fey2}
\ee
Eq. (\ref{fey2}) is
dependent on a selected measurement output such as
$E$  after a time $\tau$, for a  measuring instrument that 
incurs  an error  $E_r$ during the measurement. 
The use of the Gaussian measure, 
$w_{[E]}$=$\exp{\left\{ -{\langle(H_0-E)^2 \rangle \over \Delta E^2} \right\} }$ 
enables the effect of the measurement to be incorporated via
the  effective Hamiltonian \cite{ono1} for a two-level system
\be
\widehat H_{eff} = \widehat H_0-i{\hbar\over{\tau  E_r^2}}(\widehat H_0-E)^2
\label{measure}
\ee
The term within  $\langle ... \rangle$ in the expression for 
$w_{[E]}$ denotes the time-average for the duration $\tau$
during which  measurement  was performed. 
$E_r$,  the error incurred during the measurement of the energy, $E$,
can be taken to be a measure of the precision of the monitoring device. 
A large error made during the measurement can be viewed 
as a weak or unsharp measurement and $\widehat H_{eff} \rightarrow 
\widehat  H_0$, whereas one made with very small error can be
considered a highly precise measurement.

During a measurement process, the system evolves
as 
\be 
i \hbar {\partial \over \partial t}\ket{\psi(t)}= 
H_{eff} \ket{\psi(t)}
\ee
By expanding 
the state of the  system being measured within the unperturbed 
basis states $\ket{n}$
of the unmeasured system with Hamiltonian $\widehat H_0$ as
$|\psi(t) \rangle =\sum_n C_n(t) |n \rangle$, the coefficients
$ C_n(t)$ can be solved via the Schr\"odinger equation  in
Eq.(\ref{measure}).

\section{Appearance of Exceptional points during quantum measurements}\label{excepts}
In the ideal unmeasured state, the  Hamiltonian $\widehat H_0$ 
(see  Eq.(\ref{measure}))
of the qubit state associated with a  two-level system
 with energies $E_1$ ($E_2$) at state $\ket{0}$
($\ket{1}$)  is given by
\be
\label{qubit}
\widehat H_0
= - \hbar (\frac{\Delta \omega}{2}\, \sigma_{z} + V(t)\, \sigma_{x}),
\ee
where the Pauli matrices $\sigma_{x} = \ket{0} \bra{1}
+ \ket{1} \bra{0}$,  $\sigma_{z} = \ket{1} \bra{1}
- \ket{0} \bra{0}$ and $\Delta \omega=2(E_1+E_2)$.
 We consider that the two-level qubit is coupled
to an external   potential of magnitude   $V(t)$,
which induces transitions between the two levels. The  perturbation potential terms
are taken to be $V_{00}=V_{11}=0$ and
$V_{01} = V_{10}^\ast = V_0 e^{i \omega (t-t_0)}$ with $V_0$ as a real number.
 
For the simple model of the two-level system, the state of the  measured system,  $\ket{\psi(t)}$
evolves as
\be
 \label{evolveF}
\ket{\psi(t)} = e^{-i(E_1- i \lambda_1/4) t} C_1(t)\ket{0}
+ e^{-i(E_2- i \lambda_2/4) t} C_2(t)\ket{1}
\ee
where $\lambda_1$=$\frac{(E_1-E)^2}{2 \tau E_r^2}$ and
$\lambda_2$=$\frac{(E_2-E)^2}{2 \tau E_r^2}$ for a renormalized $E_r$.

The coefficients  $C_1(t), C_2(t)$ in Eq.(\ref{evolveF}) are  obtained using
\begin{equation}
 \left[ \begin{array}{c}
  C_1(t) \\
  C_2(t) \\   \end{array} \right] =
\left[
  \begin{array}{cc}
    \cos {\kappa}t-i \alpha_1 
&-i\alpha_2\\
    -i\alpha_2 & \cos {\kappa} t+i\alpha_1 
\\  \end{array}  \right]\
\left[ \begin{array}{c}
    C_1(0) \\   C_2(0) \\
  \end{array}
\right],
\end{equation}
where $\alpha_1$=$\cos \theta \sin{\kappa} t$, 
 $\alpha_2$=$\sin \theta \sin{\kappa} t$,
  $\cos \theta$=$\frac{q}{\kappa}$, 
$\kappa$=$\sqrt{q^2+V_0^2}$, $q$=$\frac{1}{2}(\omega-\Delta E+ i \Omega/2)$, 
$\Delta E$=$(E_2-E_1)$, and  $\Omega$=$\lambda_2$-$\lambda_1$. The  terms 
$\lambda_2$ and $\lambda_1$ as defined in the earlier paragraph
are dependent on the measurement precision, $E_r$ as well as the
energy $E$ ($E_1$ or $E_2$) to be measured. It is to be noted that 
analogous solutions for other two-level systems undergoing
dissipations have been obtained in earlier works \cite{gar,staf}.

The qubit  states  of the monitored  system therefore 
incorporate non-Hermitian terms which are functions of the measurement attributes
\bea
\nonumber
\ket{\chi_{\rm s}(t)}  & = & e^{- \lambda_t t/4} \left(\cos {\kappa}t-i \cos \theta \sin{\kappa t}\right)
 \ket{0} \; \\ \nonumber&& - i e^{- \lambda_t t/4} \sin \theta \sin{\kappa} t\ket{1}  \\ \nonumber
\ket{\chi_{\rm a}(t)}  & = & e^{- \lambda_t t/4}  \left(\cos {\kappa}t+i \cos \theta \sin{\kappa t} \right)\ket{1} \;
 \\ \nonumber && -i e^{- \lambda_t t/4} \sin \theta \sin{\kappa t} \ket{0},
\\
\label{eq:statesEigen}
\eea
where  $\lambda_t$=$\frac{\Delta E^2}{2 \tau E_r^2}$.
For measurement procedures which introduce very large 
errors,  $E_r \rightarrow \infty$, $\lambda_1$=$\lambda_2$=$\lambda_t$=$\cos \theta$=0, 
and the  qubit oscillates coherently between the
two levels with the Rabi frequency $2 \kappa=2 V_0$ as is well known in the unmeasured
system.

For a system in which the initial state at $t=0$ is 
 $\ket{1}$ and the final state 
at time $t$ is either $\ket{1}$ or $\ket{0}$,
the probability $P_{11}$ ($P_{10}$) of the 
system to be in the state  $\ket{1}$ ($\ket{0}$)
depends on the   relation between $V_0$ and $\lambda_t$.
At  the resonance frequencies, $\omega =\Delta E$, 
the Rabi frequency $2 \kappa_0= (4 V_0^2 -(\frac{\lambda_t}{2})^2)^{1/2}$, and 
$\cos\theta = -i\lambda_t/4 \kappa_0$. There are two 
 regimes, depending on the relation between $V_0$ and $\lambda_t$. The range where 
$V_0 > \frac{\lambda_t}{4}$ applies to the coherent tunneling regime where
\bea
\label{co1}
P_{11} &=&  e^{-\lambda_t t/2}
\left[\cos{\kappa_0} t- \frac{\lambda_t}{4 \kappa_0}\sin{\kappa_0} t \right]^2 
\\ \label{co2}
 P_{10} &=&   e^{-\lambda_t t/2} \frac{V_0^2}{\kappa_0^2}\sin^2{\kappa_0} t,
\eea

At $V_0 < \frac{\lambda_t}{4}$, the system undergoes incoherent tunneling
\bea
\label{inco1}
P_{11} &=&  e^{-\lambda_t t/2}\left[\cosh{\kappa_0} t- 
\frac{\lambda_t}{4 \kappa_0}\sinh{\kappa_0} t\right]^2,
 \\ \label{inco2}
P_{10} &=&   e^{-\lambda_t t/2} \frac{V_0^2}{\kappa_0^2}\sinh^2{\kappa_0} t
\eea
where as noted earlier,  $\lambda_t$=$\frac{(E_2-E_1)^2}{2 \tau E_r^2}$.
At the exceptional point, $\kappa_0 = 0, \; V_0 = \frac{\lambda_t}{4}$, and both regimes merge and 
we obtain
\bea
\label{except}
P_{11} &=&  \left(1- \frac{\lambda_t t}{4} \right)^2 e^{-\lambda_t t/2} \\
 P_{10} &=& \left(\frac{\lambda_t t}{4} \right)^2 e^{-\lambda_t t/2}
\eea
The total probabilities, $P_{11}$+$P_{10} \le 1$,
the loss of normalization is dependent on the measurement
precision, $E_r$ as expected. At the exceptional point, the population
difference, $P_{11}-P_{10}= \left(1- \frac{\lambda_t t}{2} \right) e^{-\lambda_t t/2}$,
and thus undergoes steep decay with time $t$.
The two-level system undergoing decay can be seen as a non-ideal dissipative quantum  system
due to its coupling to a multitude of decay states associated with the measurement
process.

\subsection{Exceptional points and passage times}

For the quantum measurements which manifests in Eqs. ~(\ref{co1})- ~(\ref{except}), the exceptional point 
appears at a critical measurement precision 
\bea 
E_c=\frac{\Delta E}{2 \sqrt{2 \tau V_0}}
\eea
For the specific case of $\tau=2 /V_0$ and a unit system in which $\hbar=V_0=\Delta E$=1,
we obtain  $E_c=\frac{1}{4}$, so that 
at  $E_r > \frac{1}{4}$ ($E_r < \frac{1}{4}$), the quantum system undergoes coherent (incoherent) tunneling.

\begin{figure}[htp]
  \begin{center}
{\includegraphics[width=6.55cm]{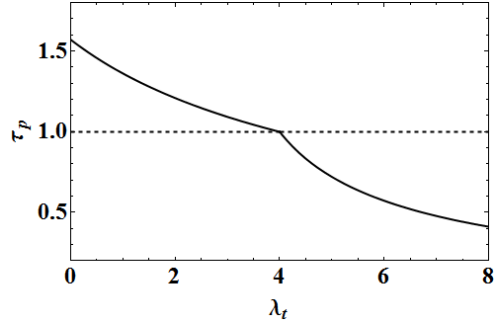}}\vspace{-1.1mm} \hspace{1.1mm}
          \end{center}
  \caption{ Passage time $\tau_p$ versus measurement precision parameter, $\lambda_t$.
A unit system in which $\hbar=V_0$=1 is used. The passage time at the exceptional point is indicated by the
dotted lines. }
 \label{pass}
\end{figure}

The passage time is defined as the smallest time $\tau_p$ taken by an 
initial state to evolve to the final state, which is generally
orthogonal to it \cite{brody1,brody2}.  Here we consider $\ket{1}$ as the  initial state,
and $\ket{0}$ as the final state, and estimate the passage time by setting
$P_{11}=0$. This  provides the passage  time, 
\be
\tau_p= ({1}/{\kappa_0})\tan^{-1} ({4 \kappa_0}/{\lambda_t})
\ee at
$V_0 > \frac{\lambda_t}{4}$, and 
for $V_0 < \frac{\lambda_t}{4}$, we obtain 
\be
\tau_p= ({1}/{\kappa_0})\tanh^{-1}({4 \kappa_0}/{\lambda_t})
\ee
At the exceptional point, 
\be
\tau_p = 4/\lambda_t
\ee
The above results are consistent with those obtained by Nesterov \cite{nestsigma}
who used a different criteria to evaluate the passage times.
For a unit system in which  $\hbar=V_0$=1, $\lambda_t$=4 at the exceptional point, and we obtain
$\tau_p$=1 as shown by the dotted lines in Fig.~\ref{pass}. 
The same figure shows the gradual decrease
in $\tau_p$ with the measurement precision at $\lambda_t < $ 4 (coherent regime) 
and continued decrease in $\tau_p$ at  $\lambda_t  >$ 4   in the incoherent tunneling
regime. We note that the units adopted yield a time duration, $\tau$=2, hence 
the passage time at the exceptional point   is half of the measurement duration. 
At the limit $\lambda_t \rightarrow 0$,
$\tau_p \rightarrow \frac{\pi}{2}$, and at $\lambda_t \rightarrow \infty$, we obtain $\tau_p \rightarrow 0$.

\section{Dynamics of quantum correlations and entanglement of observed qubits}\label{disco}

We consider a composite system consisting of the  two-level system ($s$),
the source of  the external   potential  $V(t)$ (denoted by  $r$), and the 
external monitoring device (denoted by  $d$),  with a measurement precision, $E_r$.
During the measurement procedure,
the composite state $\ket{\psi_i}=\ket{1}_s\ket{0}_r \ket{0}_d$ evolves
under the action of the Hamiltonians of Eq.~(\ref{measure}) and (\ref{qubit}) as
\be
\label{evolve}
    \ket{\psi_t}=\xi_t \ket{1}_s\ket{0}_r \ket{0}_d +\eta_t \ket{0}_s\ket{1}_r \ket{0}_d + \chi_t\ket{0}_s\ket{0}_r \ket{1}_d 
\ee
where $|\xi_t|^2=P_{11}$, $|\eta_t|^2=P_{10}$ (see Eqs. ~(\ref{co1}) to (\ref{inco2}) )
and $|\chi_t|^2=1-|\xi_t|^2-|\eta_t|^2$. We note that in the composite state $\ket{\psi_i}=\ket{1}_s\ket{0}_r \ket{0}_d$,
only a single excitation is initially present in the qubit, and accordingly only one
of the considered subsystem ($s$, $r$ or $d$)
is present in the $\ket{1}$ state. Thus in the absence of excitations in either
the system $s$ or source of external potential $r$, it is considered that the 
detector states are excited. 

In this Section, we focus on the detailed analysis of a system of two noninteracting subsystems ($s_1,s_2$)
where each subsystem is equivalent to the two-level system $s_1$  in  Eq. ~(\ref{evolve}).
Each  subsystem is  monitored by its respective observers ($d_1,d_2$), with identical characteristics such that  the  measurement precision, $E_r$ is the same for both detectors. The Hamiltonian of the total system is thus given by the sum of the Hamiltonians of the two noninteracting subsystems, $ \hat H_T= \hat  H_1 + \hat H_2$, where $H_i$ (i=$1,2$) have the form
given in Eq. ~(\ref{qubit}). 
In  order to examine the response of two initially entangled qubits which are monitored by $d_1,d_2$,
we assume  that the pair of  qubits are in an initial state of the form
\be
\label{inistate}
    \ket{\Psi_i}= \left(a \ket{0_{s_1}0_{s_2}}+b \ket{1_{s_1}1_{s_2}}\right)\ket{0_{r_1}0_{r_2}}
    \ket{{0}_{d_1}{0}_{d_2}},
\ee
where $a, b$ are complex parameters, and the source of external potential $r$ and 
detector $d$ are present in their vacuum state.   Using Eq. ~(\ref{evolve}), the composite system, $\ket{\Psi_i}$ 
evolves as
\be 
   \ket{\Psi_t}=a \ket{0_{s_1}0_{s_2}}\ket{0_{r_1}0_{r_2}}\ket{0_{d_1}0_{d_2}}+ b  \ket{\psi_t}_{1}
   \ket{\psi_t}_{2},
\ee
By tracing out  the degrees of freedom of the qubits associated with the sources of external potential $r_1,r_2$, and
detectors $d_1,d_2$, the reduced density matrix  of the bipartite two-qubit system  at time $t$ is obtained 
in the basis $\{\ket{00}, \ket{01}, \ket{10},\ket{11}\}$ as
\be
\fl
\label{tqubitm}
 {\rho}_{s_1s_2}(t)=\left(
\begin{array}{cccc}
 |a|^2+|b|^2(1-|\xi_t|^2)^2  & 0 & 0 & a b^\ast(\xi_t^\ast)^2  \\
  0 & |b|^2|\xi_t|^2(1-|\xi_t|^2)  & 0 & 0 \\
  0 & 0 &|b|^2|\xi_t|^2(1-|\xi_t|^2)  & 0 \\
   a^\ast b\xi_t^2& 0 & 0 & |b|^2|\xi_t|^4 \\
\end{array}
\right)\\
\ee
where we have assumed identical characteristics for the two detectors.
A similar analysis involving the evolution of a multipartite state influenced
by the coefficients $a,b$ have been performed  for the  excitonic qubit system, but
not in the current context of quantum measurements,  in  earlier works \cite{thilaz1,thilaz2}.

The concurrence of the  density matrix in  Eq.~(\ref{tqubitm}) is obtained as \cite{woot}
\be
\label{conc}
\mathcal{C}_{s}(t)=\mathrm{max}\left\{0,2|b||\xi_t|^2[|a|-|b|(1-|\xi_t|^2)]\right\}.
\ee
The concurrences, $\mathcal{C}_{r}(t)$ and $\mathcal{C}_{d}(t)$
associated respectively with the 
two-potential source  and two-detector density matrices at time $t$ 
are obtained by respective substitutions $\xi_t\rightarrow\eta_t$ and $\xi_t\rightarrow\chi_t$ in  Eq.~(\ref{conc}).

\subsection{Dynamics of quantum correlations near exceptional points}

While the entanglement measure such as concurrence has
useful features,  a different correlation  measure known as the quantum discord
 is more robust as it captures a  nonlocal correlations not present in the  
entanglement measure.  The quantum discord is non vanishing
 in  states which has zero entanglement, however its  evaluation 
involves lengthy optimization procedures and analytical expressions are known
 exist only in a few limiting cases. It would be interesting to investigate the 
change in the  quantum correlation ($Q$) \cite{ve1,ve2} which has similar features as the
quantum discord introduced by Olliver and Zurek \cite{zu}  in the context of the
 two-qubit system undergoing quantum measurements. In particular,
the  dynamics of non-classical correlations, especially  near the exceptional point
remains unexplored in earlier works.

The quantum correlation $Q$ is obtained by computing  the  classical correlation  $C$ in the first instance.
For the two-qubit case, $C$ is obtained \cite{ ve1,ve2} using 
$C(\rho_{s_1s_2}) \equiv \max_{\{\Pi_j\}} [S(\rho_{s_1}) -
S_{\{\Pi_j\}} (\rho_{s_1|s_2})]$, where the maximum is determined from 
the set of positive-operator-valued measurements  $\{\Pi_j\}$
in the adjacent partition $s_2$. $Q$ is  evaluated  using the
quantum mutual information  $I$, which is given by the sum
of classical correlation and quantum correlations. It is to be noted that the 
quantum correlations   is dependent  on the position of the partition on which measurements are carried out,
i.e., $Q$ is asymmetric. $Q$ becomes symmetric only for 
$S(\rho_1)= S(\rho_2)$ where $S(\rho)$ denotes the Von Neumann
entropy of state $\rho$).

Following the analytical approach in Ref.\cite{qali,qali2,wangli}, expressions for the
$C$ and  $Q$ associated with the qubit-qubit bipartite system, are obtained as
\bea
\label{disexp}
C_{s}(t) &=&Q_{s}(t)={H}(|b \xi_t|^2)\nonumber\\
&-& {H}\left (\frac12 \left [1+\sqrt{1-4|b \xi_t (1-\xi_t^2)^{1/2}|^2} \right ] \right),
\eea
where  the Shannon entropy ${H}(x)
=-x{{log}}_2x-(1-x){{log}}_2(1-x)$. 
Analogous expressions for $C_r,Q_r$ and  $C_d,Q_d$ associated respectively with the 
two-potential source  and two-detector density matrices at time $t$ 
are obtained by respective substitutions $\xi_t\rightarrow\eta_t$ and $\xi_t\rightarrow\chi_t$ in  Eq.~(\ref{disexp}).
It is to be noted that for all  bipartite systems ($s_1-s_2$, $r_1-r_2$ and $d_1-d_2$)
the classical correlations are equal to quantum correlations, so that the quantum mutual information
is twice that evaluated for $Q$ for the bipartite systems.

\begin{figure}[htp]
  \begin{center}
    \subfigure{\label{fig1a}\includegraphics[width=4.55cm]{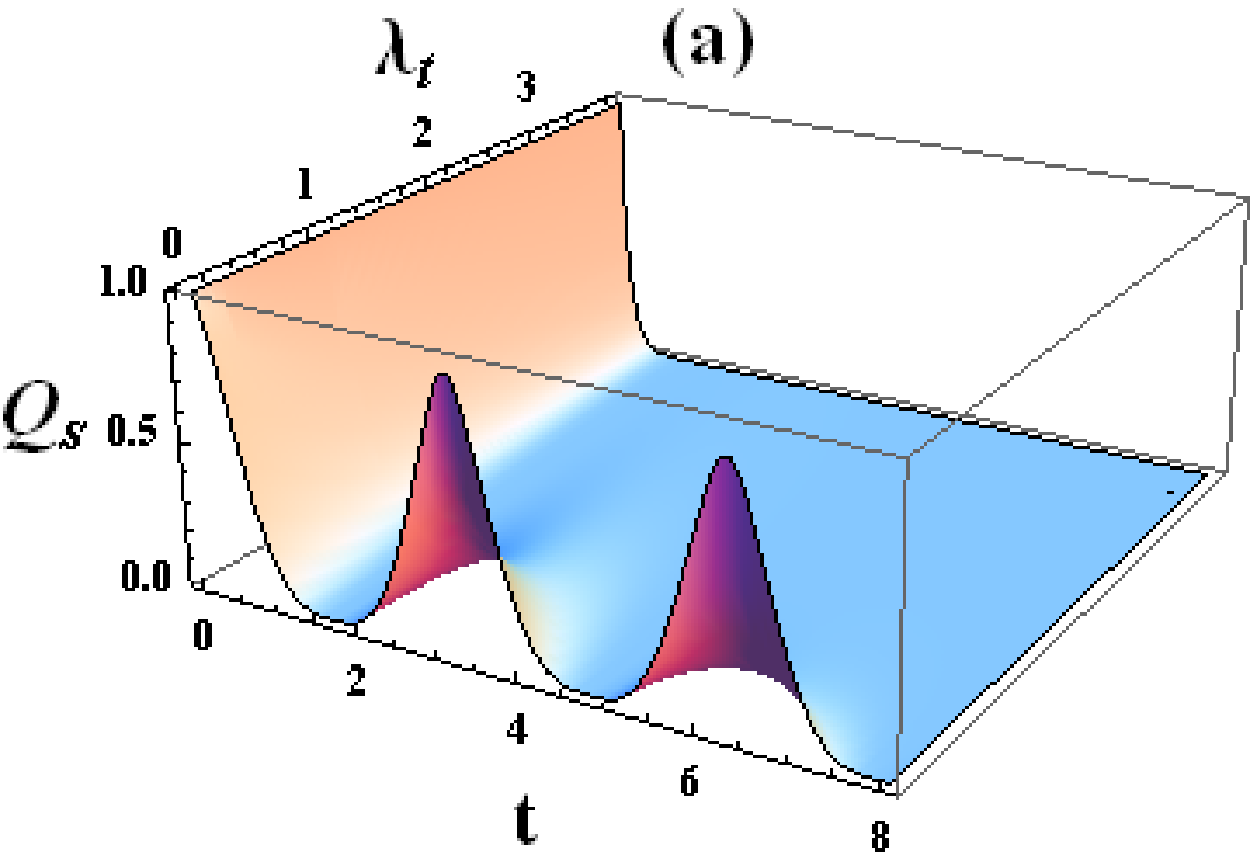}}\vspace{-1.1mm} \hspace{1.1mm}
     \subfigure{\label{fig1b}\includegraphics[width=4.55cm]{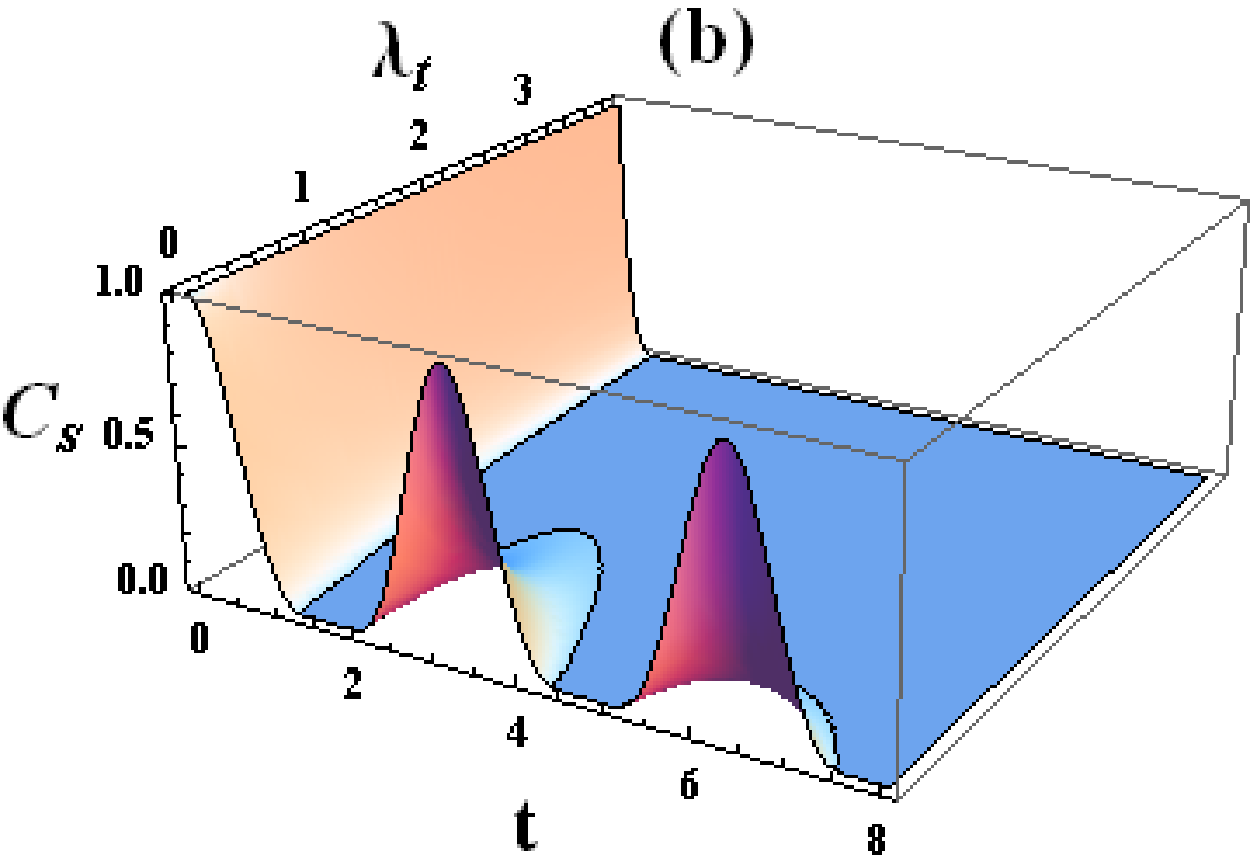}}\vspace{-1.1mm} \hspace{1.1mm}
\subfigure{\label{fig1c}\includegraphics[width=4.55cm]{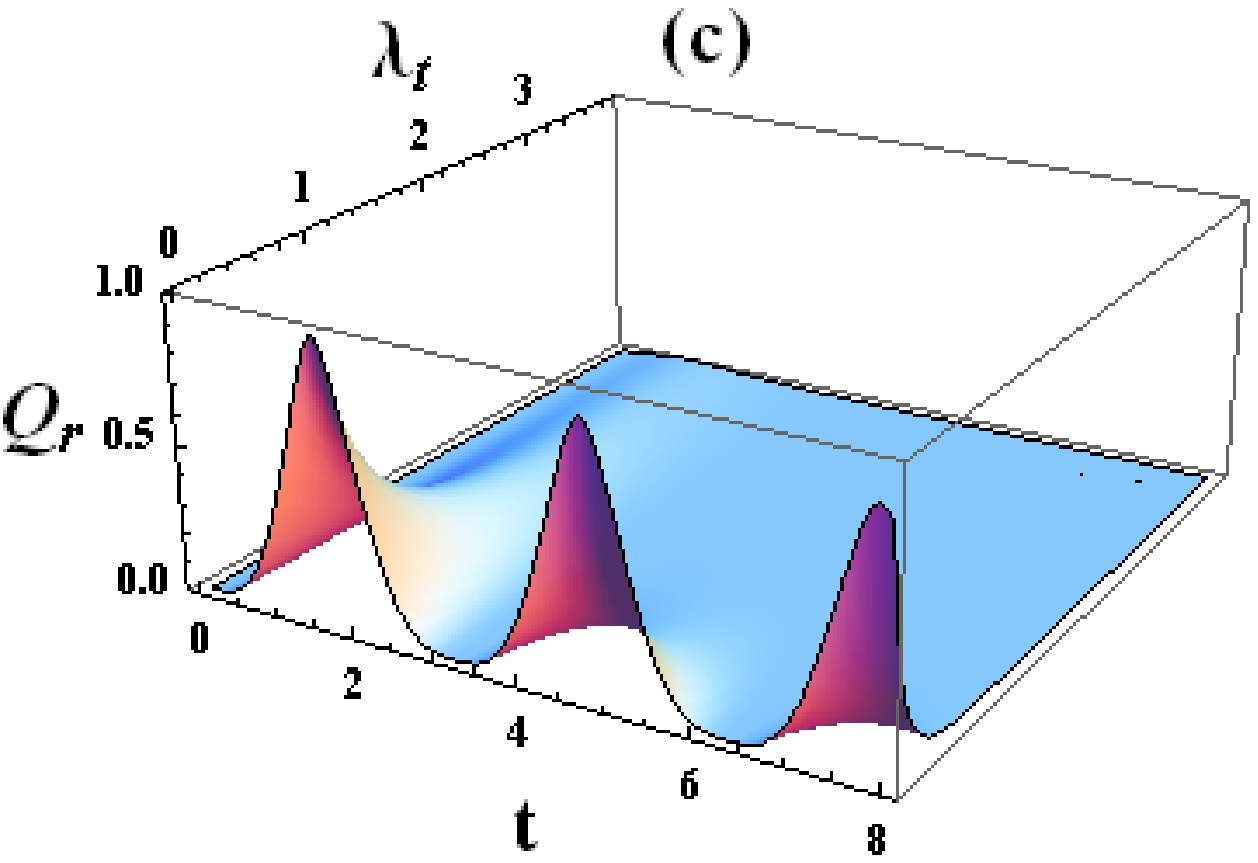}}\vspace{-1.1mm} \hspace{1.1mm}
     \subfigure{\label{fig1d}\includegraphics[width=4.55cm]{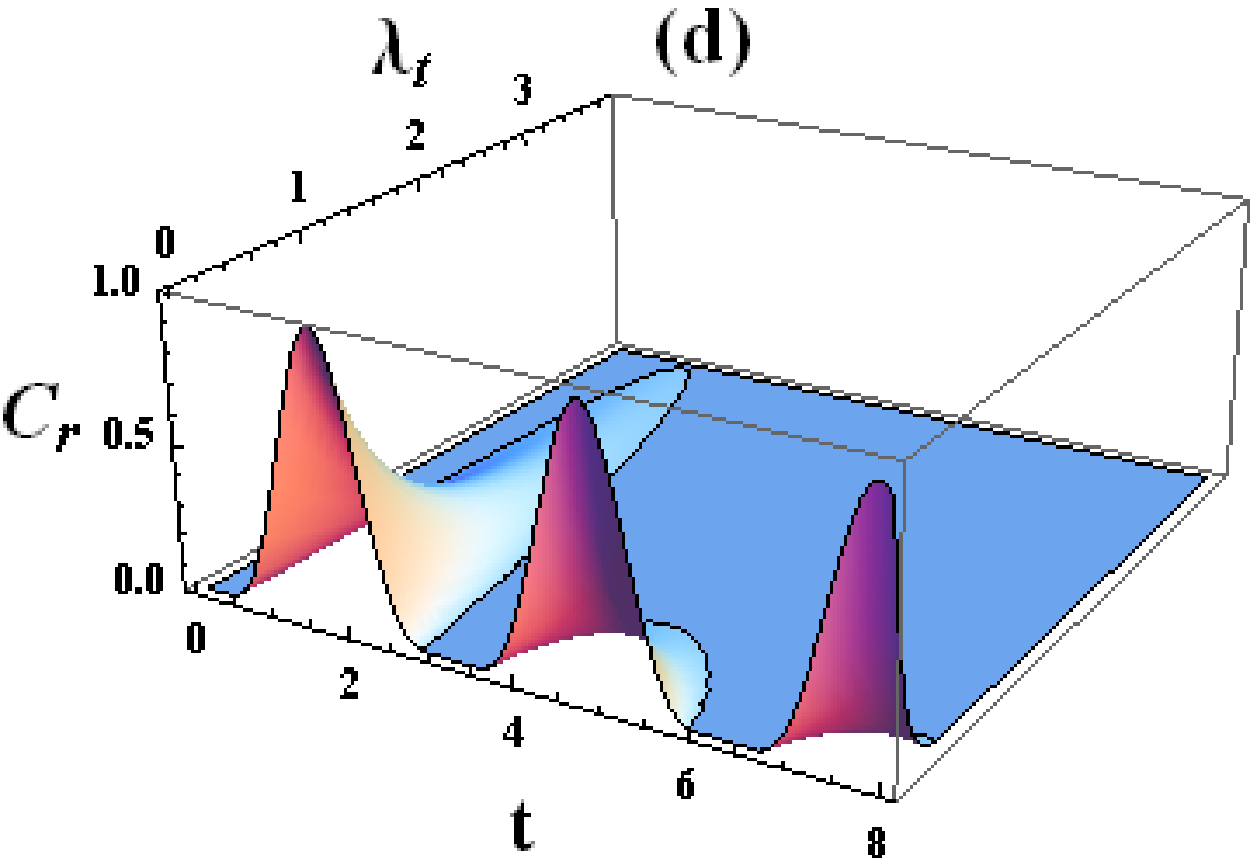}}\vspace{-1.1mm} \hspace{1.1mm}
\subfigure{\label{fig1e}\includegraphics[width=4.55cm]{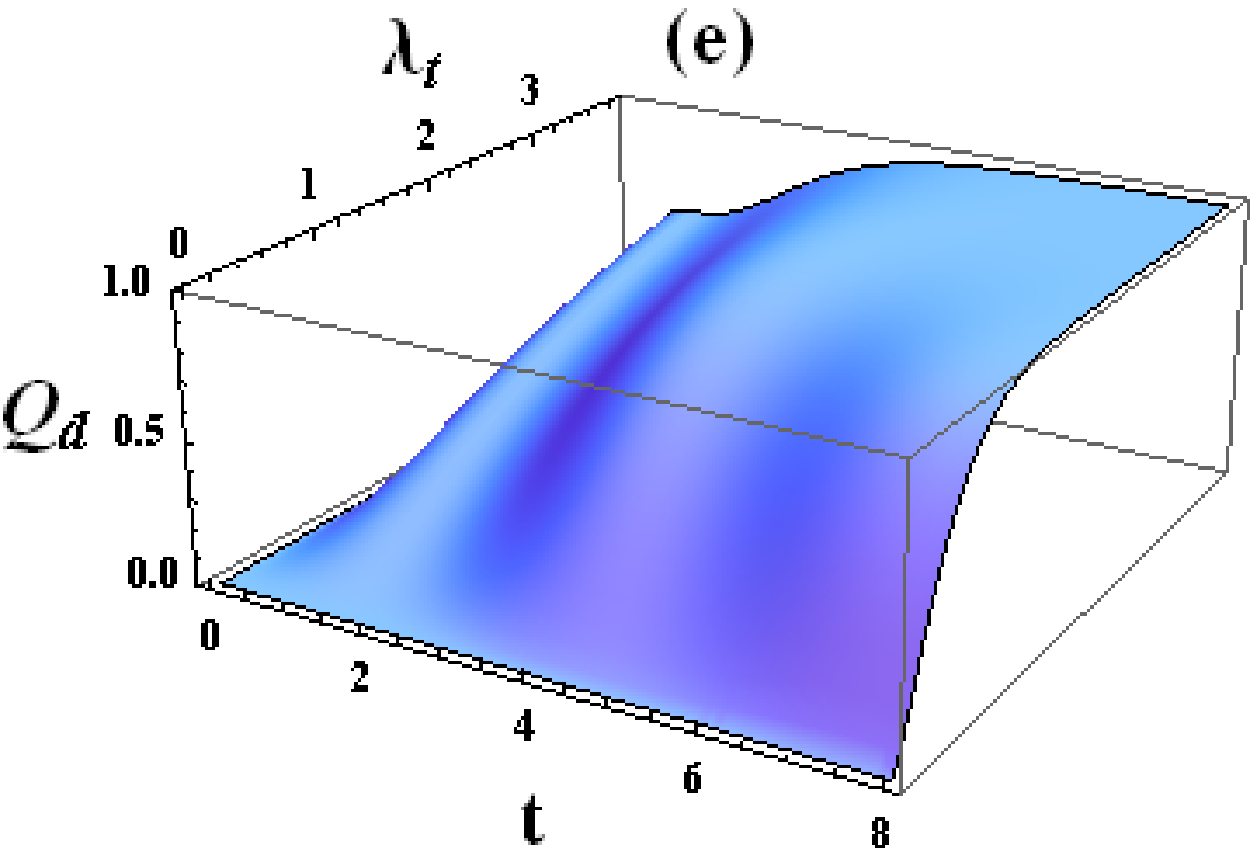}}\vspace{-1.1mm} \hspace{1.1mm}
     \subfigure{\label{fig1f}\includegraphics[width=4.55cm]{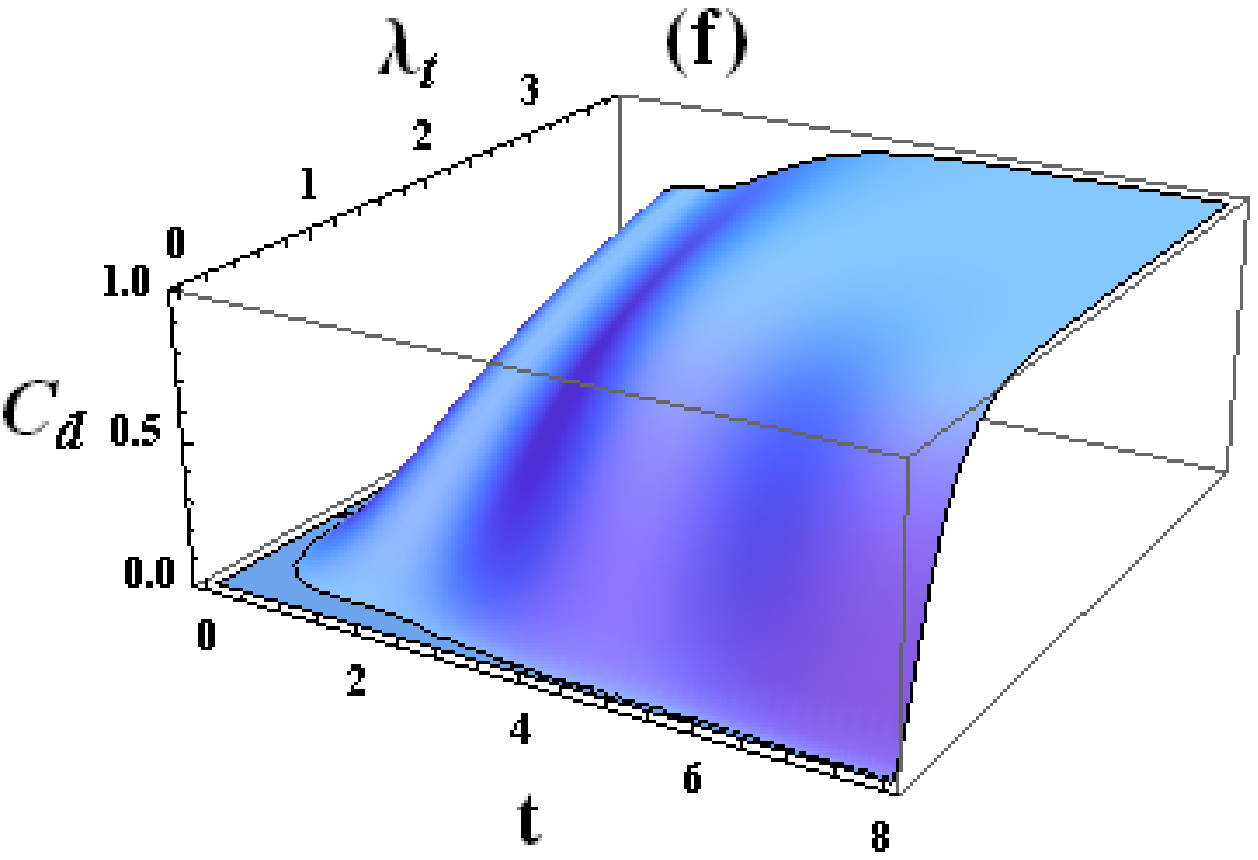}}\vspace{-1.1mm} \hspace{1.1mm}
     \end{center}
  \caption{(a)  Two-qubit quantum  correlation $Q_s$ and (b) Two-qubit concurrence $\mathcal{C}_{s}$
as function of time $t$ and measurement precision parameter $\lambda_t$
 with  initial amplitude parameter $b$=0.75. A unit system in which $\hbar=V_0=\Delta E$=1, $\tau$=8 is used,
hence 0$\le t \le$ 8 and $\lambda_t=\frac{1}{(4 E_r)^2}$
\\
(c)  Two-potential source  quantum  correlation $Q_r$ and (d) Concurrence $\mathcal{C}_{r}$
as function of time $t$ and measurement precision parameter $\lambda_t$
 with  the initial amplitude parameter   $b$=0.75 .\\
(e)  Two-detector quantum  correlation $Q_d$ and (f) Two-detector concurrence $\mathcal{C}_{d}$
as function of time $t$ and measurement precision parameter $\lambda_t$
 with  initial amplitude parameter $b$=0.75 .\\ }
 \label{disc}
\end{figure}

\begin{figure}[htp]
  \begin{center}
{\label{fig1c}\includegraphics[width=4.55cm]{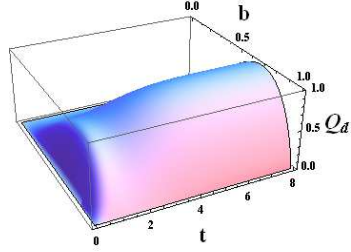}}\vspace{-1.1mm} \hspace{1.1mm}
          \end{center}
  \caption{ Two-detector  quantum  correlation $Q_d$ as function of time $t$ and
initial amplitude parameter $b$ at the exceptional point where $\lambda_t$=4.}
 \label{exceptpt}
\end{figure}

In Fig.~\ref{disc}a to \ref{disc}f, the quantum correlations 
(as a function of time $t$ and measurement precision term $\lambda_t$)
associated with the qubit-qubit,  potential-potential sources and
detector-detector bipartite systems are illustrated alongside  their corresponding concurrence measures.
For convenience, we have used  a unit system in which $\hbar=V_0$=1, so that 
time $t$ and the measurement precision parameter $\lambda_t$ are unitless as well.
The figure shows the progressive transfer of initial correlations and concurrence between qubits
to the potential-potential source qubits and then to creation of quantum correlations and entanglement between
the two detector systems with increasing $\lambda_t$.  The quantum correlations remains more robust during this transfer. As $\lambda_t$ increases, the system moves closer to the exceptional point
and the quantum correlations between the monitored (monitoring) systems  is considerably decreased (increased).
As shown in Fig.~\ref{exceptpt}, the quantum correlations in the detector-detector bipartite system 
reaches near unity values for a range of the parameter $b$ at the exceptional point and at the end of the measurement
duration.
It is apparent that the loss of information associated with the formation of exceptional points
in the qubit-qubit bipartite system correlates  to increased quantum correlations in the detector-detector
bipartite system. The results in Fig.~\ref{disc} support the view that the collapse of 
the observed state vector implies the transfer of the wave function's amplitude, and indirectly
quantum information away from the   interaction space. In this regard the non-Hermitian description of quantum measurement is consistent with increased  information transfer to the detector system in the case of
highly precise measurements.

The results thus far have been obtained for a case where  the two qubits are in a pure entangled state at $t=$0
as given in Eq. ~(\ref{inistate}). A different initial state may yield some changes in the magnitude of the
concurrence and quantum  correlations obtained here, however the overall trend in changes of these measures with 
$t$ and $\lambda_t$ is expected to be the same as illustrated in Fig.~\ref{disc}.
In summary, the results  indicate that a highly precise observer
can diminish the non-classical correlation shared between two subsystems undergoing measurements,
with the tendency to do so increasing with measurement precision.

\subsection{Relationship between the exceptional point and appearance of non-classical correlations}
The  exact relationship between the occurrence of the exceptional points
and the  quantum  correlations between the various parties (qubit, detectors) involved is not immediately clear, 
however the transfer of  non-classical quantum correlations from one subsystem  to another 
appear to underpin quantum measurements. The results obtained in this work,
 show that  as  the system moves closer to the exceptional point (small error/duration),
the quantum  correlation between the monitored (monitoring) systems  is considerably decreased (increased).
Moreover the passage time from one state to another at the
exceptional point is correlated with enhanced quantum  correlation or non-classical correlations
between the  measuring instruments. This is possibly consistent
with the fact that loss of quantum information due to the merging of eigenfunctions
at the exceptional point is translated to greater  involvement of the unobserved states of the measuring devices. 
The exact mathematical formulations underlying these non-trivial correlations and exchanges
may require greater understanding of the topological fabric space in which the exceptional points
are embedded, and such non-trivial spaces may facilitate examination 
of the degeneracies that is unique to exceptional points. For instance, Nesterov et. al. \cite{nest09,nestsigma}
has modelled the quantum evolution dynamics of the complex sphere on a one-sheeted
two-dimensional hyperboloid space  in terms of two inner parameters and obtained
results (such as distances between the final and initial states), highlighting the importance of
the geometric features of the dynamics of non-Hermitian systems. Accordingly, greater
insight on the   underlying mechanism by which entities such as entanglement and 
non-classical correlations of quantum systems are transferred from one subsystem to another
may be revealed via  use of the geometric approach developed by Anandan and Aharonov \cite{Anand}.

\section{Conclusion}\label{con}

In this work, we have examined  the  appearance of exceptional points and passage times 
in a two-level system which  admits an effective non-Hermitian description
via the   Feynman's path integral formalism and  based  on the restricted path integral approach.
The two dimensional model using an oscillatory (in time) potential as off-diagonal perturbation is
solved analytically and the time dependent transition probabilities are evaluated.
Their explicit dependence on the measurement duration time and error  
demonstrate that at resonance frequency,  there are two regimes depending on error and measurement duration: incoherent tunnelling (small error/duration) and coherent tunnelling (large error/duration); these two regimes merge at the exceptional point at which  point the passage time is noted to
 be half of the measurement duration. The entanglement dynamics of a multipartite  system consisting of the  correlated qubits, the  sources of external potential and   monitoring devices is examined using correlation
 measures such as concurrence and quantum  correlations. The results indicate that the  quantum  correlations between the monitored (monitoring) systems  is considerably decreased (increased) as  the measurement precision  nears the exceptional point. Quantum  correlations shows  greater robustness than the concurrence measure during the measurement process, which is expected due to the richness of quantum correlations intrinsically present in the former entity.

Overall the results in this study show that  quantum measurement procedures with a  select range of precision,
 can be used to transfer quantum correlations present as non-classical quantum  correlations in one system to remote systems with a  certain degree of reliability. To this end, future investigations  involving measures such as the fidelity 
may be used  to analyse the reliability range of information transfer.
The results presented here may also be extended to examine high efficiencies of energy transfer in photosynthetic  systems \cite{thilchem,thilchem2} and quantum system
relevant to quantum computation \cite{nel}. However the explicit link between the passage time
 and appearance of non-classical correlations  needs further study,  possibly 
requiring  geometric approaches as proposed in Ref. \cite{Anand}.
Nevertheless, the appearance of singularities such as the exceptional points
in  highly precise measurements, as predicted in this work,
  may provide useful guidelines to their detection 
in  experimental  studies involving quantum measurement in optics and nanostructure systems 
\cite{bus,brakha,hall, tit}. Accordingly the  results obtained here may facilitate
 the testing ground of pseudo-Hermitian quantum mechanics using appropriate
quantum measurement procedures  in nanostructured devices.

\section{Acknowledgments}
The author gratefully acknowledges  useful comments from
 the anonymous referees.  This research was undertaken on the NCI National Facility in 
Canberra, Australia, which is supported by the
Australian Commonwealth Government.


\section{References}
\begin{thebibliography}{9}

\bibitem{sense}
C.  M. Bender, Reports on Progress in Physics {\bf 70}, 947 (2007).
 
\bibitem{benderB}
C. M. Bender and S. Boettcher, Phys. Rev. Lett . {\bf 80}, 5243 (1998).

\bibitem{rotter}
I. Rotter, J. Phys. A: Math. Theor. {\bf 42}, 153001 (2009).


\bibitem{nest09}
A. I. Nesterov, Physics Letters A {\bf 373}, 3629 (2009).


\bibitem{bend}
 C. M. Bender, Rep. Prog. Phys. {\bf 70}, 947 (2007).
\bibitem{fring}
P. E. Assis and A. Fring, J. Phys. A: Math. Theor. {\bf 41}, 244002 (2008).
\bibitem{ali}
 A. Mostafazadeh,  Phys. Rev. Lett.  {\bf 99} 130502 (2007).
\bibitem{mosta}
A. Mostafazadeh, Int. J. Geom. Meth. Mod. Phys. {\bf 7}, 1191 (2010).
\bibitem{bendJ}
C. M. Bender, D. C. Brody, H. F. Jones, and B. K. Meister, Phys.
Rev. Lett. {\bf 98}, 040403 (2007).

\bibitem{heissN}
W. D. Heiss, Eur. Phys. J. D {\bf 29}, 429 (2004); Czech. J. Phys.
{\bf 54}, 1091 (2004).

\bibitem{hern}
E.Hernandez, A. Jauregui and A. Mondragon
Phys. Rev. E {\bf 84}, 046209 (2011).

\bibitem{muss}
Z. H. Musslimani, K. G. Makris, R. El-Ganainy, and D. N.
Christodoulides, Phys.Rev.Lett. {\bf 100}, 030402 (2008).

\bibitem{jin}
L. Jin and Z. Song, Phys. Rev. A {\bf 84}, 042116 (2011).

\bibitem{chem}
O. Atabeck and R. Lefebvre, J. Phys. Chem. A {\bf 114}, 3031 (2010).

\bibitem{berry}
M. V. Berry and M. R. Dennis, Proc. R. Soc. London A {\bf 459},
1261 (2003).

\bibitem{micro2011}
B. Dietz, H. L. Harney, O. N. Kirillov, M. Miski-Oglu, 
A. Richter, and F. Sch\"afer, Phys. Rev. Lett. {\bf  106}, 150403 (2011).


\bibitem{exptex}
M. Philipp, P. von Brentano, G. Pascovici, and A. Richter, Phys.
Rev. E  {\bf 62}, 1922 (2000).

\bibitem{Goet}
A. Goetschy and S. E. Skipetrov
Phys. Rev. E {\bf 84}, 011150 (2011).

\bibitem{lee}
S.-B. Lee, J. Yang, S. Moon, S.-Y. Lee, J.-B. Shim, S. W. Kim,
J.-H. Lee, and K. An, Phys. Rev. Lett. {\bf 103}, 134101 (2009).


\bibitem{thilchem}
A. Thilagam, J. Chem. Phys. {\bf 136}, 065104 (2012).


\bibitem{Heiss}
W. D. Heiss, Phys. Rev. E {\bf 61}, 929 (2000).


\bibitem{Anand}
J. Anandan and Y. Aharonov, 
 Phys. Rev. Lett. {\bf 65}, 1697 (1990).



\bibitem{nestsigma}
A.I. Nesterov, SIGMA {\bf 5} 069  (2009).

\bibitem{von}
J. von Neumann, ``Mathematical Foundations of Quantum Mechanics,''
Princeton University Press, Princeton, (1955).

\bibitem{zeh}
H. D. Zeh, Found. Phys. {\bf 1}, 69 (1970).


\bibitem{ob}
A. G. Kofman, G. Kurizki, Nature {\bf 405}, 546 (2000);
Phys. Rev.A {\bf 54}, 3750(R)(1996).


\bibitem{Misra}
B. Misra and  E. C. G. Sudarchan, J. Math. Phys. {\bf 18}, 758 (1977).

\bibitem{bus}
 P. Busch, P. J. Lahti, and P. Mittelstaedt, 
``The Quantum Theory of Measurement,'' Springer-Verlag, Berlin, (1991).

\bibitem{wise}
H. M. Wiseman, Quantum Semiclass. Opt. {\bf 8} 205 (1996)  



\bibitem{It}
W.M. Itano, D. J. Heinzen, J. J. Bollinger and D. J. Wineland,
Phys. Rev.A {\bf 41}, 2295 (1990).


\bibitem{FacJ}
P. Facchi and  S. Pascazio, J. Phys. A: Math. Theor. {\bf 41} 493001 (2008);
P. Facchi  and  S. Pascazio, Phys. Rev. Lett. {\bf 89}, 080401 (2002).


\bibitem{brakha} 
V. B. Braginsky and F. Ya. Khalili, ``Quantum Measurement", K. S. Thorne 
editor (Cambridge University Press, Cambridge) (1992), and references cited 
therein. 

\bibitem{menbk} 
M. B. Mensky, ``Continuous Quantum Measurements and Path-Integrals" 
(Institute of Physics Publishers, Bristol and Philadelphia) (1993). 

\bibitem{zuro}
W. H. Zurek, Phys. Today {\bf 44 (10)}, 36 (1991). 

\bibitem{schloss}
M. Schlosshauer, ``Decoherence and the Quantum-to-Classical
  Transition", Springer-Verlag, (2008).


\bibitem{joh}
J. Kofler and C. Brukner, Phys. Rev. Lett. {\bf 101}, 090403 (2008).



\bibitem{gar}
 J. C. Garrison and E. M. Wright, Phys. Lett. A, {\bf 128} 177 (1988).


\bibitem{Feyn1} R. P. Feynman and A. R. Hibbs,
``Quantum Mechanics and Path Integrals", (McGraw-Hill, New York) (1965).


\bibitem{Feyn2}
 R. P. Feynman,  Rev. Mod. Phys. {\bf 20}, 367 (1948).

\bibitem{mensky} M. B. Mensky, Phys. Rev. D {\bf 20}, 384 (1979);
Sov. Phys. JETP {\bf 50}, 667  (1979).


\bibitem{menprl}
 M. B. Mensky, R. Onofrio, and C. Presilla, 
Phys. Rev. Lett. {\bf 70},  2825 (1993).


\bibitem{men}
M. B. Mensky, R. Onofrio, and C. Presilla, 
Phys. Lett. A {\bf 161}, 236 (1991).


\bibitem{ono1} 
R. Onofrio, C. Presilla, and U. Tambini, 
Phys. Lett. A {\bf 183}, 135 (1993). 




\bibitem{staf}
C. A.  Stafford and B. R. Barrett, Phys. Rev. C {\bf 60}, 051305 (1999).


\bibitem{brody1}
 D. C. Brody,  J. Phys. A: Math. Gen. {\bf 36} 5587 (2003).


\bibitem{brody2}
D. C. Brody and D. W.  Hook, J. Phys. A: Math. Gen. {\bf 39} L167 (2006).


\bibitem{thilaz1}
A. Thilagam, J. Phys. A: Math. Theor. {\bf 43}, 155301 (2010).

\bibitem{thilaz2}
A. Thilagam , J. Phys. A: Math. Theor. {\bf 42}, 335301 (2009).

\bibitem{woot}
V. Coffman, J. Kundu and W. K. Wootters, Phys.
Rev. A {\bf 61}, 052306 (2000).




\bibitem{ve1} L. Henderson and V. Vedral, J. Phys. A {\bf 34}, 6899 (2001).


\bibitem{ve2}  V. Vedral, Phys. Rev. Lett. {\bf 90}, 050401 (2003).



\bibitem{zu} H. Ollivier and W. H. Zurek, Phys. Rev. Lett. {\bf 88}, 017901 (2001).


\bibitem{qali} M. Ali, A. R. P. Rau, and G. Alber, Phys. Rev. A {\bf 81}, 042105 (2010).

\bibitem{qali2} M. Ali, A. R. P. Rau, and G. Alber, Phys. Rev. A {\bf 82}, 
069902(E) (2010).


\bibitem{wangli} 
C.Z. Wang, C.X. Li, L.Y. Nie and J.F. Li, J. Phys. B: At.
Mol. Opt. Phys. 44,  015503 (2011).


\bibitem{thilchem2}
A. Thilagam, J. Chem. Phys. {\bf 136}, 175104  (2012).

\bibitem{nel}
M. A. Nielsen and I.L.  Chuang,
{\it Quantum Computation and Quantum Information}
(Cambridge University Press, Cambridge, U.K., 2000).

\bibitem{hall}
M. A. Hall, J. B. Altepeter and P. Kumar 
Optics Express {\bf 17}, 14558(2009).

\bibitem{tit}
W. Tittel, J. Brendel, H. Zbinden, and N. Gisin, Phys. Rev. Lett.
{\bf 81}, 3563 (1998).

\end{thebibliography}
\end{document}